\documentclass[b5paper,twoside]{jpconf}  
\usepackage{graphicx}

\newcommand{\beq}{\begin{equation}}
\newcommand{\eeq}{\end{equation}}
\newcommand{\beqn}{\begin{eqnarray}}
\newcommand{\eeqn}{\end{eqnarray}}

\begin{document}

\title[Transit Observations]
{Investigating Close-in Exoplanets through 
        Transit Observations}  

\author[Jiang et al.]{Ing-Guey Jiang$^1$, Li-Chin Yeh$^2$, 
Parijat Thakur$^3$, Ping Chien$^1$, Yi-Ling Lin$^1$, 
Yu-Ting Wu$^1$, Hong-Yu Chen$^1$,
Zhao Sun$^{4,5}$, Jianghui Ji$^4$}

\address{$^1$Department of Physics and Institute of Astronomy,\\
National Tsing-Hua University, Hsin-Chu, Taiwan
}
\address{$^2$Department of Applied Mathematics,\\
National Hsinchu University of Education, Hsin-Chu, Taiwan
}
\address{$^3$Department of Pure \& Applied Physics 
and Institute of Technology,
Guru Ghasidas Vishwavidyalaya (A Central University),
Bilaspur (C.G.) - 495 009, India
}
\address{$^4$Purple Mountain Observatory, Chinese Academy of Sciences, 
Nanjing 210008, China
}
\address{$^5$Graduate School of Chinese Academy of Sciences, 
Beijing 100049, China
}

\ead{jiang@phys.nthu.edu.tw} 

\begin{abstract}
Through the international collaborators, 
we recently established a network of existing and working 
meter-class telescopes to look 
for planetary transit events. As a first step, we 
focus on the TrES3 system, and conclude that there 
could be some level of 
transit timing variations.
\end{abstract}

\section{Introduction}

The development of the research in extra-solar planetary systems 
has been successful 
continuously for nearly two decades.  
In addition to the steadily increasing number of newly detected systems,
many new aspects of discoveries also make  
this field extremely 
exciting. It is the Doppler-shift method to make many extra-solar
planetary systems known to us. In addition, the method of transit,
micro-lensing, and direct imaging also produce fruitful results. 
In particular, 
more than 100 extra-solar planets (exoplanets) have been 
found to transit their host stars.  Moreover, Kepler space telescope
is revealing many interesting results recently.
For example, one of the greatest accomplishment is the discovery 
of a system with 6 planets. These results really demonstrate the
state-of-art power of the transit technique.

Among these detected planetary systems, TrES3 (O'Donovan et al. 2007)
attracts many attentions due to its strong transit signal and 
short period comparing with other known systems detected before 2007. 
For example,  Sozzetti et al. (2009) presented a nice work 
which combines both spectroscopic and photometric 
observations of TrES3 to obtain the best models 
for the host star and planet. 
With an aim to study the possible transit timing variations,
Gibson et al. (2009) did nine transit observations.

On the other hand, Fressin et al. (2010) even use Spitzer space telescope
to monitor TrES3 during its secondary eclipse.
The most important constraint from their results is to 
show that orbital eccentricity is almost zero, consistent with a circular 
orbit assumption.
Furthermore, 
the 10.4-m Gran Telescopio Canarias 
(currently the world's largest optical telescope) has obtained 
extremely high-precision narrow-band transit data for the 
TrES3 system (Colon et al. 2010).
There is almost no deviation for data on the light curve, 
so this data could give a strong constraint on orbital parameters 
of exoplanet TrES3b.
 
\section{Observations and Data Reduction}

In this project, we monitor the TrES3 system 
by the 0.8 meter telescope of Tenagra Observatory
in Arizona, America. In May and June 2010, five runs of 
transit observations were done by our group successfully.  
By the differential photometry performed 
through the software IRAF, five light curves are obtained.

In addition to these five light curves, many more data are 
taken from Exoplanet Transit Database (ETD), which is an internet site 
supported by Czech Astronomical Society (Poddany et al. 2010).
The TrES3b transit data available in Colon et al. (2010),
which employed a 10.4 meter telescope, 
is also included into our analysis here. 
ETD classifies their data into 5 classes, from Quality 1 to 5 
based on the flux deviation of light curves.
In this analysis, seven Quality 1, fourteen Quality 2, 
and two Quality 3 data are taken to be used.
Table 1 lists all ETD data employed in this paper.

\begin{table}[h]
\begin{center}
\begin{tabular}{llll}
\hline
 Epoch &    Filter  &   Quality  &        Author \& Reference  \\ \hline 
   885 &    Clear  &      1    &        Valerio Bozza, Gaetano Scarpetta \\
   876 &    Clear &      1    &        Garlitz J. \\
   837 &    Clear &      1    &       Tanya Dax, Stacy Irwin, Karissa Haire \\
   695 &     R    &      1    &         Scarmato T.\\
   401 &     Clear &      1    &         Gajdos S, Jaksoval I.\\
   336 &    Clear   &    1    &        Gajdos S, Jaksoval I. \\
   306 &    V      &     1    &        Cao Ch. \\
   872 &     Clear  &     2    &       Sergison D. \\
   834 &     Clear  &    2    &       Shadic S \\
   709 &     Clear   &   2    &       Shadick S.\\
   654 &      R      &   2    &       Hose K. \\
   598 &     r sloan  &   2    &      Vanhuysse M. \\
   595 &     V      &    2    &       Westall K. \\
   411 &      Clear &    2    &       Gajdos S, Jaksova I. \\
   389 &      V     &    2    &      Vander Haagen G.\\
   372 &      R     &    2    &      Veres P,Kornos L., Toth J.\\
   349 &      I     &    2    &      Naves \\
   348 &      V     &    2    &       Cao Ch. \\
   333 &      R     &    2    &      Gregorio \\
   294 &     Clear  &    2    &       Trnka J \\
    86 &       I    &     2   &      Moon \\
   720 &   R   &   3  &       Sun Z. \\
   284 &   R   &   3  &       Naves\\
\hline
\end{tabular}
\caption[Table ETD]{The list of employed ETD data}
\end{center}
\end{table}

\section{The Light Curves} 

The Transit Analysis Package (TAP) which was developed and described in 
Gazak et al. (2011) is used for our 
light-curve analysis. The Markov Chain Monte Carlo (MCMC)
technique and the model of Mandel \& Agol (2002) are employed to fit
the light curves in TAP.
A recent paper by Fulton et al. (2011) provides a good description
for other details of techniques in TAP.
Because TAP uses simple analytical models of Mandel \& Agol (2002),
we will only employ one set of transit data to get orbital parameters 
through TAP, and thus allows minor potential orbital differences  
between epochs. This process will also allow other orbital parameters
to be changed in order to get the best values of mid transit times.

Before using TAP,
the target's flux is normalized so that 
the values are close to the unity during out-of-transit. 
Moreover, as described in Eastman et al. (2010), 
BJD is used for the time stamps in light curves.

These light-curve data are then filled into TAP to start MCMC chains.
To start a MCMC chain in TAP, we need to set the initial values of 
following parameters: 
orbital period $P$,
orbital inclination $i$,
the semi-major axis $a$ (in the unit of stellar radius $R_{\ast}$),
the planet's radius $R_{\rm p}$ (in the unit of stellar radius),
the mid transit time $T_{\rm m }$, 
the linear limb darkening coefficient  $u_1$,
the quadratic limb darkening coefficient  $u_2$,
orbital eccentricity $e$, 
and longitude of periastron $\omega$.
Once the above initial values are set, 
one can choose any one of the above to be 
(1) completely fixed, (2) completely free to vary,
or (3) varying following a Gaussian function, i.e. Gaussian prior,
during the MCMC chain in TAP.
The details are listed in Table 2.
 
\begin{table}[h]
\begin{center}
\begin{tabular}{lll} 
\hline
parameter & initial value & during MCMC chains \\
\hline  
$P$(days)& 1.30618581&a Gaussian prior with $\sigma=0.0002 $ \\
$i$(deg)  & 81.85         &  free       \\
$a$/$R_{\ast}$  & 5.926  &    free             \\
$R_{\rm p}$/$R_{\ast}$&0.1655&a Gaussian prior with $\sigma=0.02$\\
$T_{\rm m }$ & set-by-eye &   free    \\
 $u_1$&Claret (2000,2004)&a Gaussian prior with $\sigma=0.05$\\
 $u_2$&Claret (2000,2004)&a Gaussian prior with $\sigma=0.05$\\ 
$e$              &   0.0    &  fixed  \\
 ${\bar \omega}$ &   0.0   &  fixed    \\
\hline
\end{tabular}
\caption[The setting of parameters]{
The setting of parameters}
\end{center}
\end{table} 

\section{The Results and Concluding Remarks}

\begin{figure}[h]
\begin{minipage}{14pc}
\includegraphics[width=14pc]{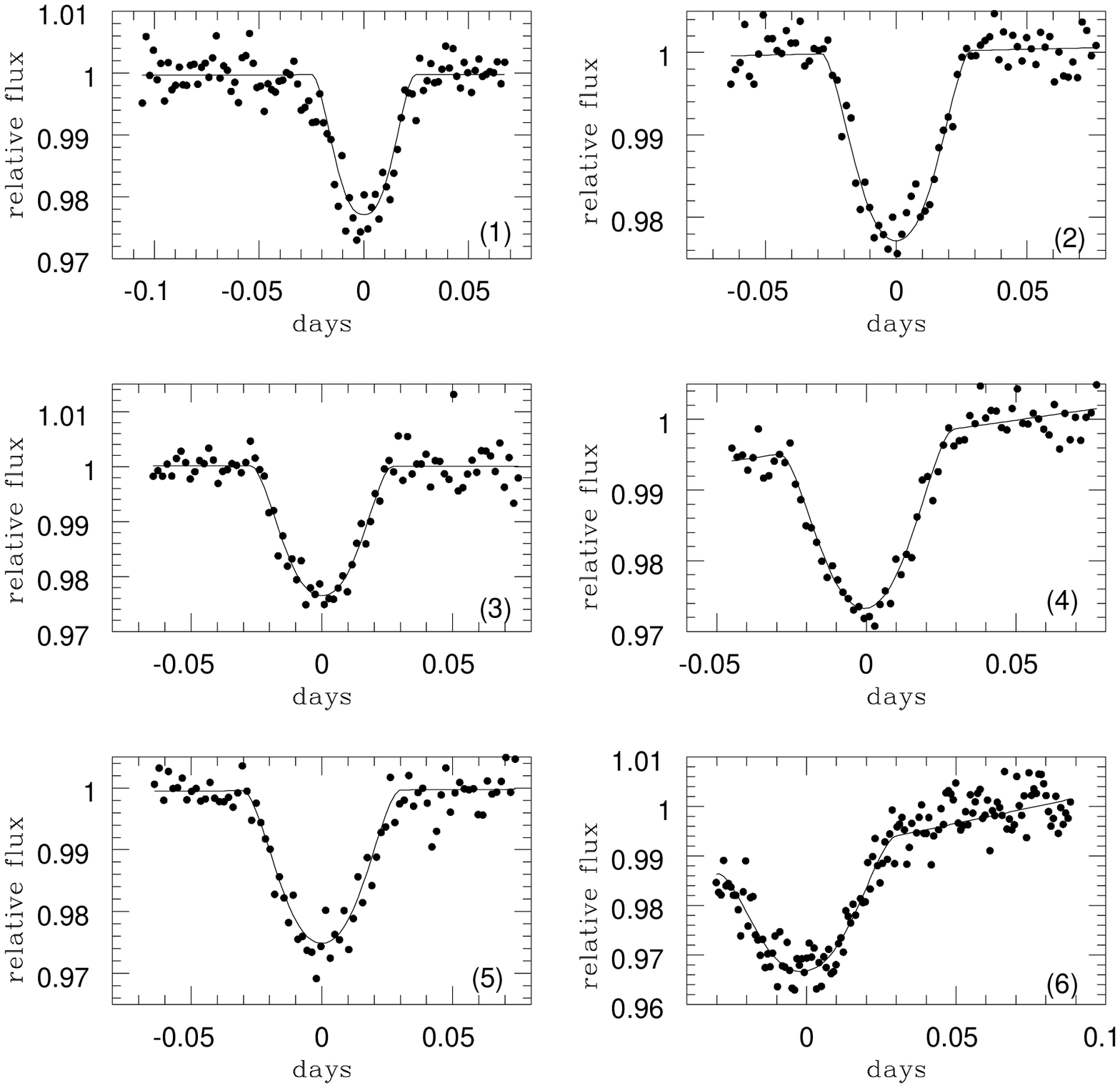}
\caption{\label{label}Transit light curves: points are the data and
curves are models. Panel (1)-(5) 
are our own Tenagra data, and Panel (6) 
is Sun Z.'s data in ETD (Epoch 720).}
\end{minipage}\hspace{2pc}%
\begin{minipage}{14pc}
\includegraphics[width=14pc]{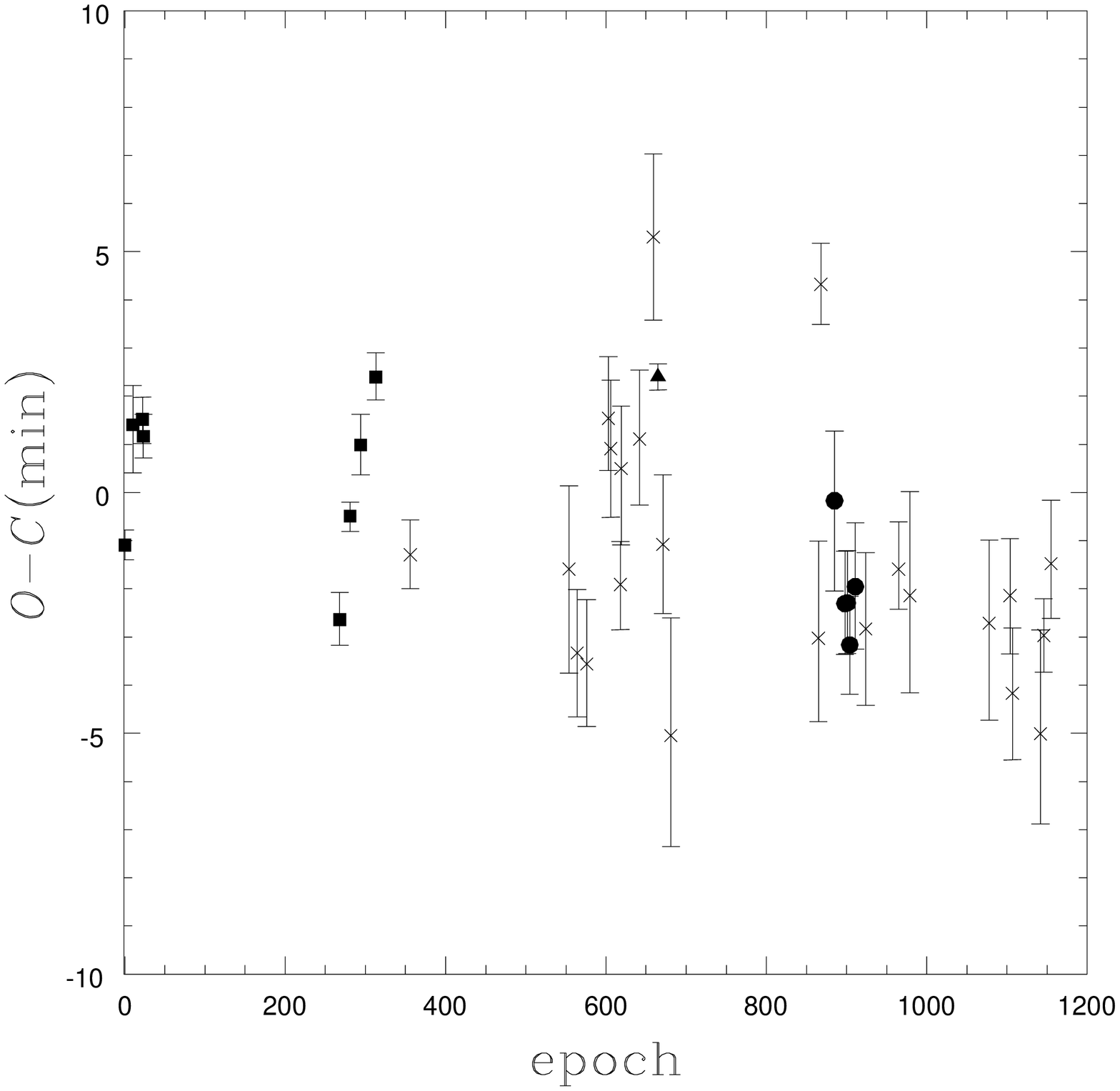}
\caption{\label{label}$O-C$ diagram. 
The full circles are for our Tenagra data, squares are for the data from 
Sozzetti et al (2009), the full triangle is for 10.4 meter GTC data,
and crosses are for ETD data. }
\end{minipage} 
\end{figure}
 
Fig. 1 shows our light curves and best fitting models. 
Fig. 2 presents the $O-C$ diagram, in which the differences between
the observed mid transit time, $O$, and the calculated mid transit
time of a simple star-planet two-body system, $C$, are plotted.
To produce this diagram, we first have to assign the epoch $E$ to 
each transit event. The one with smallest 
mid transit time, $T_{\rm m}$, would be given epoch $E=0$, 
and the epochs of all the rest data can be assigned according to their 
$T_{\rm m}$ easily.
Using a linear function, $T^C_{\rm m}= E P + T_0$, where 
$E$ is treated as a variable and $P$, $T_0$ are constants,
to do chi-square fitting
on all the available observational  data set of $(E, T_{\rm m})$,
we can get the best $P$ and $T_0$.
The plot $T_{\rm m}-T^C_{\rm m}$ as a function of epoch $E$,
in which the best $P$ and $T_0$ are used for $T^C_{\rm m}$,
is the $O-C$ diagram.

Note that we do not allow period $P$ to change freely,
but only vary slightly around the value given in 
Sozzetti et al. (2009),
and that variation of 0.0002 days is only about 0.3 min.
The variation of mid transit times shown in Fig. 2 are obviously
much larger. In the  $O-C$ diagram,
the transit timing variation (TTV) is about 3 to 4 min.
Based on this result, if we use a straight line of zero variation 
to fit those data in $O-C$ diagram, we obtain a reduced $\chi^2$ to be 8.77.
Thus, the assumption with no TTV provides a bad fit to the data, and
some level of TTV probably exists.


\section*{References}
\begin{thebibliography}{4} 

\bibitem{Cl2000} 
Claret, A. 2000, A\&A, 363, 1081

\bibitem{Cl2004} 
Claret, A. 2004, A\&A, 428, 1001

\bibitem{C} Colon, K. D. et al. 2010, MNRAS, 408, 1494

\bibitem{E} Eastman, J., Siverd, R., Gaudi, B. S. 2010,
PASP, 122, 935

\bibitem{F} Fressin, F. et al. 2010, ApJ, 711, 374

\bibitem{Fu} Fulton, B. J. et al. 2011, AJ, 142, 84

\bibitem{Ga} 
Gazak, J. Z., Johnson, J. A., Tonry, J., 
Eastman, J., Mann, A. W., Agol, E.
2011, arXiv:1102.1036

\bibitem{G} Gibson, N. P. et al. 2009, ApJ, 700, 1078


\bibitem{M} Mandel, K., Agol, E. 2002, ApJ, 580, L171

\bibitem{O}
O'Donovan, F. T. et al. 2007, ApJ, 663, L37

\bibitem{P}
Poddany, S., Brat, L., Pejcha, O. 2010,
New Astronomy, 15, 297

\bibitem{So}
Sozzetti, A. et al. 2009, ApJ, 691, 1145 


\end{thebibliography}
\end{document}